# Anomaly Detection in Emails using Machine Learning and Header Information

Craig Beaman
Craig.Beaman@unb.ca
Canadian Institute for Cybersecurity
Faculty of Computer Science
University of New Brunswick
Fredericton, Canada

Haruna Isah
h.isah@unb.ca
Canadian Institute for Cybersecurity
Faculty of Computer Science
University of New Brunswick
Fredericton, Canada

*Abstract*—Anomalies in emails such as phishing and spam present major security risks such as the loss of privacy, money, and brand reputation to both individuals and organizations. Previous studies on email anomaly detection relied on a single type of anomaly and the analysis of the email body and subject content. A drawback of this approach is that it takes into account the written language of the email content. To overcome this deficit, this study conducted feature extraction and selection on email header datasets and leveraged both multi and one-class anomaly detection approaches. Experimental analysis results obtained demonstrate that email header information only is enough to reliably detect spam and phishing emails. Supervised learning algorithms such as Random Forest, SVM, MLP, KNN, and their stacked ensembles were found to be very successful, achieving high accuracy scores of 97% for phishing and 99% for spam emails. One-class classification with One-Class SVM achieved accuracy scores of 87% and 89% with spam and phishing emails, respectively. Real-world email filtering applications will benefit from the use of only the header information in terms of resources utilization and efficiency.

## I. INTRODUCTION

The use of email offers many advantages to both individuals and companies in that it is easy to use, cheap, and convenient. It allows for messages and any number of attachments to be quickly sent between individuals or groups of people. The sent messages are also often stored permanently on email servers, ensuring that a record of the messages is always available. However, the ease of use and convenience that email offers does not come without a price, for instance, spam and malicious emails can easily be sent to thousands of people. Such malicious emails can be thought of as anomalies when compared to normal emails. This work categorizes emails into three types: ham, spam, and phishing emails. Ham emails refer to 'normal' emails without any malicious intent, that is to say, any email that is not a spam or phishing email. Spam emails refer to unsolicited bulk emails [1]. Phishing emails pretend to come from a trusted source with the goal of causing harm to the individual or organization [2].

Email structure follows the guidelines defined in RFC 5322 and consists of two major components: the header and the body. The headers consist of a number of different fields, where each field has both a name and a value that are separated by a colon and a space. Examples of email header fields are "From", "To", "Subject", "Date", "Message-ID", and "Received". Most header fields can be spoofed in some cases, but some are harder to spoof than others [3]. For example, the Received fields tend to be difficult to spoof [3]. Most of the header information is created by the sending agent of the email, but some information, such as information related to routing (e.g., Received field), is added to the header as the email is routed through various email servers. The email body comes after the header fields and contains the message text to be read by the recipient and attachments [4]. In addition to the headers defined in RFC 5322, there are also X-headers, which are an extended set of email headers that are defined by individual email services, such as Gmail or Outlook.

Email anomalies can vary in form and intent, from simple advertisements to spear-phishing campaigns where users are tricked into giving up personal information or tricked into clicking on malicious attachments. However, anomalies could include any email that differs significantly from normal behaviour or patterns. Considering email anomalies not as just spam or phishing emails hence has an advantage in that this approach ought to generalize better to more types of potential threats. For example, if an employee's email account was taken over by an attacker, and that employee's account then starts to send large numbers of malicious emails to coworkers, this could be seen as an anomaly by the system. However, such behaviour may not be picked up by traditional spam or phishing detection systems since it comes from an inside source to the company by a trusted individual [5].

Being able to identify anomalies in email datasets is thus extremely important from a security perspective. For example, in 2018 US-based organizations lost a reported $2.7 billion from email-based attacks [5]. It is estimated that over 150 million phishing emails are sent out each year, with around sixteen million of those making it through filters and landing in unsuspecting employee and personal inboxes. Successful phishing attacks often take close to a year to identify and fully contain the damages [6]. From an individual perspective, these types of crimes can lead to identity theft, financial loss, and the installation of malware [7, 8]. A recent study has shown that as much as 88% of malware is distributed through email [9]. This all highlights a clear need for effective countermeasures.

In terms of research trends, there have been substantial amounts of research into various forms of anomaly detection using a wide variety of features and various email datasets. However, most of the research efforts heavily focused on supervised learning using the email body and subject content. Additionally, these works have been found to focus on a single type of anomaly (i.e., either spam or phishing emails), and have not looked into detecting both together in one system. Furthermore, it has been observed that very little research has focused on unsupervised learning and on systems that focus on email header content.

This study leverages machine learning to detect anomalies such as spam and phishing in emails using only the header information. Perhaps one advantage of focusing on email headers over email body content is that header features can be easily made language agnostic (i.e., the written language of the email does not matter). For example, many email header features simply compare if two values are equal or not; hence such features do not need to understand the meaning or semantics of those strings. Many features that focus on email body content do so by analyzing the frequencies of 'spam-like' words. However, if the language shifts from say English to French, and the model was only trained on English, then this would lead to issues. Using only header information avoids such problems. Additionally, using header fields avoids the issues of image spam, which is when spam or phishing messages are placed in images to make analysis more difficult [10]. In particular, feature extraction and selection will be performed on a dataset of emails. The resulting features will only contain email header information; neither body nor subject content will be considered in this work. From there, various anomaly detection algorithms will be utilized and compared and conclusions will be drawn about the effectiveness of detecting anomalies strictly from using email header data.

The rest of the paper is outlined as follows: Section II presents an overview of anomaly detection techniques and a survey on spam and phishing email detection approaches. Section III goes through the experimental design. Section IV goes through the testing strategies and results. Section V discusses and compares the results, and Sections VI and VII discuss future research directions and conclusions, respectively.

## II. LITERATURE REVIEW

### A. Anomaly Detection

*1) Definition of Anomalies:* Anomalies are irregular patterns or behaviours in data that significantly differ from expected or normal behaviour. Outliers and novelties are similar terms related to anomalies. They both denote rare examples in a dataset. While outliers represent examples that come from the expected model but are simply unlikely or rare outcomes, novelties appear when the underlying normal model for the data shifts over time leading to the emergence of new examples. [11]. An analogy using baseball cards as examples in the dataset would go as follows: an anomaly would be a hockey card that was thrown in, an outlier could be a 1st edition Babe Ruth card (rare, but still a baseball card), and a novelty could be a newly released type of holographic baseball card. In practice, however, the definitions are not so important, as they can all be detected in similar ways. They mainly differ in how they are handled once they are found. For example, anomalies are often of great interest, such as for fraud detection. Outliers meanwhile are typically seen as unwanted noise and are removed from the dataset.

Anomalies are identified based upon the context that is being considered; what is defined as normal and what is an anomaly can change [12]. For example, an individual having a height of over six feet may be seen as an anomaly in Vietnam, but in Denmark, this may be seen as normal. In general, anomaly detection is an automated approach used in the identification of rare events or observations which raise suspicions or differ significantly from the majority of the data. For instance, an online credit card purchase could be flagged anomalous if the purchase is far much more than normal and at a strange location or time of the day.

*2) Types of Anomalies:* Anomalies can be grouped under point, contextual, low-level sensory, and high-level semantic anomalies [11]. Point anomalies refer to individual data points which differ significantly from the vast majority of other data instances. An example of this could be the high score in a competitive video game where this score is ten times higher than that of all other players. Contextual or conditional anomalies exist when a particular data point is anomalous relative to neighboring data points; neighboring here refers to data points that have some sense of order, such as points ordered in time or space. An example of this could be in time series data of power output, where the output jumps up by 100 volts for a second or two. Low-level sensory and high-level semantic anomalies, mostly concern deep learning anomaly detection.

*3) Anomaly Detection Approaches:* Anomaly detection can apply to any of the following learning settings: supervised, semi-supervised, and unsupervised. The unsupervised approach tends to be the most popular type while the supervised approach is the least common. Supervised anomaly detection essentially reduces to a binary classification problem, where there is one class for 'normal' examples and another for 'anomalous' examples [11]. For semi and unsupervised learning, however, different techniques are required. Examples of anomaly detection algorithms include One-Class SVM, k-means clustering, kNN, and Gaussian Mixture models [11].

Anomaly detection algorithms can be classified in a few different ways based on the nature of the approach. These are statistical, distance-based, density-based, clustering-based, graph-based, ensemble-based, and learning-based classes; the advantages and disadvantages of each of these can be found in [13]. Other studies that surveyed anomaly detection techniques include [11, 14].

### B. Anomaly Detection in Emails

*1) Non-ML Based Approaches:* There are numerous approaches for countering spam emails that do not rely upon

machine learning. Some of these approaches have been outlined in [15, 16]. These techniques share one common disadvantage, they are static in nature. Meanwhile, those sending malicious emails can always change their approach to evade current countermeasures, and hence human intervention is often required for such approaches to have any success in the long-term [17]. An automatic, robust, and effective detection mechanism that is capable of identifying email anomalies is needed. Such a detection mechanism is something that machine learning-based approaches can offer. An important benefit of non-machine learning approaches is that they provide a solid foundation for sensible features that can be used by machine learning models.

*2) ML Based Approaches:* There are three different approaches to machine learning, these are supervised, semi-supervised, and unsupervised. While there are several semi-supervised and unsupervised machine learning methods that have been leveraged for spam classification, supervised learning has been by far the most popular approach to email classification in recent years. Notable supervised learning methods include the study by Gallo et al. [18] which noted that although current spam filters are quite good, they do not catch all of the threats. Using an enterprise dataset consisting of about 12,000 labeled either as malicious or benign emails, the authors extracted and utilized a wide array of features to train and evaluate a few different learning algorithms. These include Gaussian Naive Bayes, Decision Tree, Linear SVM, RBF SVM, MLP Neural Networks, and Random Forest. The study found that using just eight features allowed for a performance close to that of the full set, but with better training speeds. The best results were found when using the Random Forest and RBF SVM classifiers, achieving AUC scores of 0.981 and 0.966 for Random Forest and SVM on the full feature set, respectively. The F1 scores for the full feature set tests were 0.892 for Random Forest and 0.866 for RBF SVM.

Cidon et al. [5] looked into the problem of Business Email Compromise (BEC) and employee impersonation and developed a supervised approach called BEC-Guard to detect BEC and employee impersonation. BEC and employee impersonation are difficult problems to overcome compared to traditional spam email filtering. Such attacks are hard to detect as they often do not contain malware and are personalized to the recipient of the email. The approach by Cidon et al. was twofold, first, they analyze an email to see if it is impersonating an employee or not, and if it's found to be likely impersonating someone, then the email is further analyzed for its content and any links contained in the email body. The impersonation classifier uses Random Forest and features derived from the sender, receiver, CC, and BCC fields. The derived features also include whether or not the sender's email uses the same domain as the company, whether or not the reply-to address equals the sender's address, among others. If the impersonation classifier classifies an email as an impersonation attempt, then the email is forwarded to the second line of defense; the content classifier. The content classifier analyzes both the text in the email body, as well as any links that appear in the message. The text classifier pre-processes the text before applying a term frequency-inverse document frequency (TF-IDF) score to words appearing in the email dataset. These scores are used as features for a kNN classifier. The link classifier uses Random Forest and its features include things such as the link domain's popularity and the date the link domain was registered. By under-sampling a dataset of about 7,000 BEC and impersonated emails, the method achieved a precision of 98.2% with a false positive rate of less than one in five million emails.

Karim et al. [15] surveyed the literature of AI-assisted spam email detection methods. They observed ANN, DNN, Naive Bayes, Decision Tree, Random Forest, Logistic Regression, SVM, Adaboost, and kNN are the most frequently used classifiers. A few studies were found to have used hybrid techniques. Commonly used unsupervised approaches include K-Means and Self-Organizing Map (SOM). They noted that few studies have used IP source and destination, or SMTP envelop information. Instead, most studies have focused on SMTP header and email body content. In terms of popular trends, it was noted that supervised learning is a very popular learning approach, and that Naive Bayes and SVM were among the most popular algorithms. However, only a little research has focused on semi-supervised approaches.

A comprehensive review of ML-based approaches in spam classification can be found in [15]. Of the works that have used semi-supervised learning, ANN, Naive Bayes, SVM, Random Forest, KNN, and others have been used. For unsupervised learning, Self-organizing maps (SOM), K-means, Expectation–maximization (EM), and KNN have been used.

*C. Email Header Features*

This section is focused on features relating to email header information. Because the goal of this study is to explore the use of only the email header information for classifying anomalous emails, features relating to email body content were not considered. Additionally, the Subject header field was also left out, as features relating to the Subject header involve similar analysis to that of email body content. Table I lists the five noted categories of email header features found in the literature and also provides some examples.

For domain matching features in the comparison-based features category, some sources use partial matching instead of strict matching. The edit distance is an example of a similarity measure, where a small edit distance may indicate phishing [18]. If there are more than two domains to compare, some options include taking the best match, an average of the matching scores, returning whether or not at least one matches exactly, or creating a new feature for each matching score.

III. DESIGN

As discussed in the previous sections, many techniques have been developed by researchers and practitioners to ensure that only valid emails are delivered to the end-user. These techniques have been designed to measure the risk score of each incoming email and filter out anomalous instances to curtail the dangers posed by phishing, email-borne malware, and ransomware to users. The ML-based techniques first train a classifier in an offline setting with both anomalous emails such as spam and legitimate emails often called ham. Once trained, the classifier is used to filter incoming emails, where emails classified as anomalous or spam are either dropped or placed in a spam folder. The classifier is dynamically updated using feedback learning.

TABLE I: The five identified categories of email header features identified from literature. Note that the examples are not the full set of features, rather these are just a small selected subset of the total features which serve to better illustrate each category. The noted references contain further examples for each respective category.

| Category | Description | Examples | References |
|---|---|---|---|
| Missing Field Based Features | Whether or not a particular header field is present in an email or not. | **Missing Message-ID field**: If the Message-ID field is present in the email header, then a value of 0 is given for the feature, otherwise the value of 1 is given. | [19–23] |
| Counting Based Features | Using the count of something as a feature. | **Number of hops**: The number of Received header fields.<br>**Number of recipients**: The number of email addresses listed in the To, Cc, or From header fields. | [18–20, 22, 24–26] |
| External Lookup Based Features | Features that cannot be calculated offline, such as features requiring DNS queries. | **IP Address Legality**: Checks if the sending host's IP (which can be found in the first Received field) address exists.<br>**Number of Blacklisted SMTP Servers**: Counts how many, if any, of the SMTP servers are blacklisted (based on information in the Received fields). | [5, 18–21, 24] |
| Header Value Based Features | Extracting the field values directly and using them as a feature, or checking the field values for some condition. | **Time Zone**: The time zone listed in the Date field.<br>**Content-Type Content**: Whether or not the content type in the Content-Type header field is 'text/html'.<br>**Message-ID Domain**: Uses the domain address of the Message-ID header field as a feature. | [5, 22, 24, 24–28] |
| Comparison Based Features | Comparing two or more header values to one another. | **Similarity between Domain Addresses**: Checks whether or not the domain names found in certain header fields (e.g., Message-ID + From; From + 'from' part of first Received; From + Return-Path, etc.) are the same or similar.<br>**Received Field Consecutive Domains**: Looks at whether or not consecutive Received fields all have a matching 'from' and 'by' domain. For example, the first Received field's 'by' domain would be matched to the second Received fields 'from' domain, and so on for the other Received fields. If one of the pairs does not match, then a value of 0 is assigned for this feature, otherwise, 1. | [5, 18–20, 22–24, 29] |

In section II, it was found that little research has focused solely on email headers for email classification. Due to this deficit, this problem has been tackled in this section, where both traditional multi-class and one-class classification approaches were used to detect various anomalies in emails using email header data alone.

Figure 1 shows the architecture of the proposed email anomaly detection system, which is based only on email header information. Although this architecture shows only the offline component, it can easily be adapted for online filtering of incoming emails. The following subsections describe each of the sub-components in the architecture.

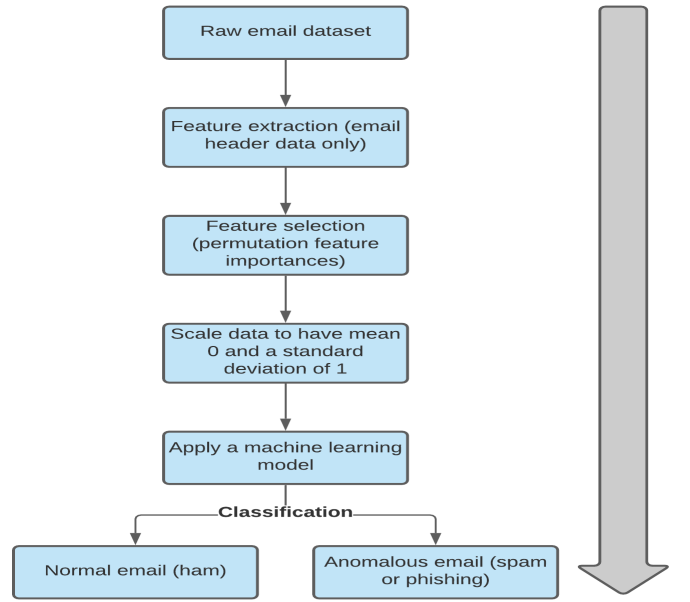

Fig. 1: The architecture of the proposed email anomaly detection system.

### A. Dataset Description

Two datasets were used in this study. The first dataset is from 2007 and consists of roughly 75,000 spam/ham emails containing the full email headers and body text. The emails were all collected from the same server over the course of about three months. The dataset can be downloaded from https://plg.uwaterloo.ca/~gvcormac/treccorpus07/about.html. This dataset will be referred to as dataset A for the remainder of the paper.

The second dataset was constructed by appending together a few smaller datasets of phishing emails and consists of 1288 total emails. These smaller datasets were collected from a single server between the years 2017 and 2020. The emails in this dataset appear in their entirety, containing both the full email headers and body text. The phishing datasets can be downloaded in their original form from https://monkey.org/~jose/phishing/. This dataset will be referred to as dataset B for the remainder of the paper.

*1) Dataset Extraction:* Both datasets A and B could not be used directly and some processing had to be done first to extract the header data. After processing, both datasets consisted of rows and columns, where each row represents a single email, and each column represents a single header field. The processing done here should not be confused with feature extraction, which is discussed later on in section III-B.

Dataset A, as downloaded, consists of thousands of separate email files. The labels of each email, either ham or spam, can be found by using a few index files that come with the dataset. Python code found at https://github.com/ParthavDesai/m.l._for_email_headers/blob/main/src/data/make_dataset.py was used to extract the emails and assign the correct label to each. This code was modified to extract a different subset of features and to further analyze the dataset. The extracted dataset, along with modified code used for creating the dataset and viewing the most common headers, can be found at https://github.com/kregg34/EmailHeaderAnomalyDetection. In total, over 1000 unique email header fields were found in dataset A, however, many of those fields appeared very infrequently. For example, only 79 of the header fields appear in more than 1000 emails (there are 75,419 emails total). Due to this, only the most common 50 email header fields were extracted for later use.

Dataset B was extracted using a modified version of the code used to extract dataset A. This code can be downloaded from https://github.com/kregg34/EmailHeaderAnomalyDetection. The same 50 header fields were extracted from dataset B as were extracted from dataset A. It is worth pointing out that the distribution of which header fields are appearing the most often in emails is very different between datasets A and B.

## B. Feature Extraction

Before feature extraction began, the datasets were analyzed to get an idea of the formats of the various header fields. From this, it was found that many fields consist of multiple parts and that the formatting often varies. For example, the Received field often consists of a few different lines, where one line might state where the email came from, another might say what server received it, and another might mention the date and time of arrival. Furthermore, the particular format's even within a single type of header field often were inconsistent. This is due to some fields having no set format listed in the relevant RFC's, and hence different email servers or clients use different formats. Another challenge with feature extraction was that many header fields appeared in only a small number of total emails; this resulted in many missing values for some features. There are also some complete outliers in terms of email header formatting, where the header field value was incorrectly formatted. This all made the pre-processing and feature extraction challenging.

To aid with parsing of the Received fields, a modified version of the code at https://github.com/Te-k/pyreceived/blob/master/pyreceived/parser.py was used. Two main approaches were taken to deal with missing values. These approaches were ordinal and one-hot encoding. Some other standard approaches to dealing with missing data, such as dropping examples with missing data or replacing missing values with an average or the most common value, could not be used; missing values could not be dropped as this would result in large amounts of data being lost, and replacing missing values by an average or most common value was not possible due to the uniqueness of many fields (e.g., many fields contained unique values for almost every example, such as the listed email addresses in the 'Reply-To' or 'From' fields).

The extracted features were based on the features surveyed in the literature, some of which are listed in Table I. However, some features, like those in the 'External look up' category, were not extracted owing to the old age of dataset A; the emails in dataset A are well over ten years old, and hence information regarding the various domains and IP addresses is likely no longer applicable for classification. Also, note that some features were dropped before testing since they contained just a single value or a very strong majority towards one value. In total, 94 features were extracted.

## C. Feature Importance and Selection

Feature importance is the process of determining what features are important to a classification model. The determination and removal of unimportant features is known as feature selection. Once the important features are identified, the less important features can be dropped without a large impact on the classification performance. Feature selection can improve accuracy, speed up training times, and reduce over-fitting [30].

There exist various feature selection algorithms, those commonly used in the domain of email classification include Permutation Feature Importance and Embedded methods. In Permutation Feature Importance, one feature at a time will have its values randomly shuffled, and the performance of the learning model is compared to a baseline value. If there is a large decrease in performance after shuffling a particular feature, then it is likely that the feature is important. Embedded methods, however, use some built-in feature selection methods of some learning models such as LASSO and RIDGE regression [31].

Because of its versatility, permutation feature importance was used in this study on dataset A using three learning models: Random Forest, SVM, and MLP. Permutation feature importance was also performed before one-class testing with an OC-SVM model, in which the top 30 features were selected.

The feature selection for dataset B was done manually. Many of the features were engineered specifically with dataset A in mind, and so applying them to dataset B did not make sense. For example, the time zone feature was created by setting the value to a 0 if the time zone of the email matched the most common time zone in dataset A, or 1 otherwise. The utilized features were the ones that check if domain addresses match or not between different header fields. These were chosen as it is believed that these may generalize well to all email datasets, since these features simply compare values, and do not use anything specific to one dataset or another.

### D. Implementation and Modeling

The Python machine learning library, Scikit-learn, was used extensively for this work. Some preprocessing of the datasets was performed using scikit-learn's 'StandardScaler' class. A variety of scikit-learn's learning algorithms, which included Random Forest, SVM, OC-SVM, MLP, KNN, were utilized. Scikit-learn's 'GridSearchCV' class was used for hyperparameter tuning and 10 fold cross-validation. Test and training sets were created using scikit-learn's 'train_test_split' class. The sets of best hyperparameters were used during later testing. The full source code can be found at https://github.com/kregg34/EmailHeaderAnomalyDetection.

## IV. TESTING AND RESULTS

The testing was conducted in the following four phases (1) Ham and Spam Binary Classification (2) Ham and Phishing Binary Classification (3) Ham and Spam One-class Classification and (4) Ham and Phishing One-class Classification. In phases 1 and 2, training was done by passing both ham and either spam or phishing emails to the learning model. In phases 3 and 4, only ham emails were passed to the model during training. For all phases, testing consisted of passing a balanced set of both ham and either spam or phishing emails. Note that the spam and ham emails all belong to dataset A, while the phishing emails all belong to dataset B.

### A. Binary Classification Approaches

The results for the binary (ham and spam) classification using several supervised learning algorithms are shown in Table III. The results show that Random Forest, MLP, SVM, and KNN perform the best.

The results for the binary (ham and phishing email) classification are shown in Table III. Note that a balanced dataset was used which consisted of half phishing emails and half ham emails.

Some tests were also done using a stacking approach of heterogeneous classifiers. This was done as stacking can improve performance compared to using individual classifiers [32]. The results of this testing for ham and spam emails are shown in Table IV. The results for ham and phishing testing are shown in Table V. Stacking was found to perform slightly better than individual classifiers.

### B. One-Class Classification Approaches

One-class classification was carried out using a OC-SVM. This approach is effective for (i) imbalanced datasets where only very few examples or none of the minority class are available and (ii) datasets where there is no coherent structure to separate the classes that could be learned by a supervised algorithm. Hyperparameter testing and ten-fold cross-validation was used. The results from testing are shown in Table VI.

### C. Deployment

In practice, anomaly detection in emails will require filtering an incoming email into a predefined class such as spam or phishing. As noted at the beginning of this section, the pipeline involves (i) the extraction and processing of relevant data (ii) the training and evaluation of a model (iii) saving the trained model in a specific or universal format, and (iv) the integration of the saved model in an application for the classification of all incoming emails. Feedback from the results is then used to optimize the model as feedback is received in an online setting. Online testing was not done in this work, however.

## V. DISCUSSION

### A. Comparisons with Other Works

There have been considerable research efforts on the topic of anomaly detection in email datasets. However, most of the effort was heavily focused on binary supervised learning using email body and subject content. Content analysis has several limitations, one of which is that it takes into account the written language that makes up the email body content. Furthermore, these research efforts are limited to a single type of anomaly in emails (spam or phishing) and have not looked into detecting both in a single pipeline. This study addressed these deficits by leveraging both binary and one-class classification approaches to detect multiple anomalies in emails using the header data only. The experimental results in the previous section demonstrate that the header information alone is sufficient for the effective detection of anomalies in emails.

The binary classification results from this study were able to achieve accuracy scores of roughly 99% for spam emails and 97% for phishing emails. The one-class classification results achieved accuracy values of around 89% for spam emails and 87% for phishing. These results either quite closely match existing studies or improve upon their results. The studies that have utilized dataset A and supervised learning include [33–36]. The studies that have utilized dataset B include [37–40]. A summary of the comparison of results is shown in Table VII.

Only a few studies used unsupervised anomaly detection along with dataset A. For instance, [41] utilized vector space model to analyze email body text and were able to achieve a weighted accuracy of 74.35%, along with a false negative rate of 42.93% and a false-positive rate of 8.36%. So far, no studies were found to use dataset B and unsupervised learning, as such direct comparisons could not be made. There are, however, other studies that have used unsupervised learning on other datasets for spam detection, these include [42–48]. Overall, the results achieved in this study compare well and with some performance improvements over those studies.

TABLE II: The results of ham and spam classification leveraging supervised learning with each algorithm using the best performing set of hyperparameters. The features correspond to the top features from permutation feature importance testing. The top three best results for each metric is highlighted in bold. All values except AUC are percentages.

| Algorithm | Accuracy | F1 | Recall | Precision | AUC |
|---|---|---|---|---|---|
| Logistic Regression | 96.6269 | 97.4817 | 98.0381 | 96.9315 | 0.9593 |
| Support Vector Machine | 99.4802 | **99.6095** | 99.5487 | 99.6704 | 0.9945 |
| Gradient Boosted Regression Trees | 99.4608 | 99.5953 | 99.6416 | 99.5491 | 0.9937 |
| Multilayer Perceptron Neural Network | **99.5651** | 99.5651 | **99.6602** | **99.6866** | **0.9952** |
| Naive Bayes (Gaussian) | 92.1577 | 94.2418 | 96.3735 | 92.2024 | 0.9006 |
| Random Forest | **99.6871** | **99.7652** | **99.8460** | **99.6849** | **0.9961** |
| Decision Tree | 99.3423 | 99.5052 | 99.3124 | **99.6988** | 0.9936 |
| K-Nearest Neighbors | **99.6111** | **99.7081** | **99.7372** | 99.6790 | **0.9955** |
| AdaBoost (Tree Stumps) | 96.0347 | 97.0552 | 98.1283 | 96.0052 | 0.9500 |

TABLE III: The results of ham and phishing classification using supervised learning algorithms with each algorithm using the best performing set of hyperparameters. The features correspond to domain matching features only. The top result for each metric is highlighted in bold. All values except AUC are percentages.

| Algorithm | Accuracy | F1 | Recall | Precision | AUC |
|---|---|---|---|---|---|
| Logistic Regression | 97.4806 | 97.3196 | 99.1597 | 95.5466 | 0.9760 |
| Support Vector Machine | **97.8682** | **97.7226** | 99.1597 | **96.3265** | **0.9796** |
| Gradient Boosted Regression Trees | **97.8682** | **97.7226** | 99.1597 | **96.3265** | **0.9796** |
| Multilayer Perceptron Neural Network | **97.8682** | **97.7226** | 99.1597 | **96.3265** | **0.9796** |
| Naive Bayes (Gaussian) | 90.3101 | 90.4943 | **100.000** | 82.6389 | 0.9101 |
| Random Forest | 97.6744 | 97.5207 | 99.1597 | 95.9350 | 0.9778 |
| Decision Tree | 97.4806 | 97.3196 | 99.1597 | 95.5466 | 0.9760 |
| K-Nearest Neighbors | 97.6744 | 97.5207 | 99.1597 | 95.9350 | 0.9778 |
| AdaBoost (Tree Stumps) | 97.6744 | 97.5207 | 99.1597 | 95.9350 | 0.9778 |

TABLE IV: The results of the stacked supervised learning algorithms with ham and spam emails. Base learners were chosen from the top performing classifiers of Table III. The meta-classifier was chosen to be Logistic Regression for all tests. All values except AUC are percentages.

| Base Learners | Accuracy | F1 | Recall | Precision | AUC |
|---|---|---|---|---|---|
| RF, MLP, kNN | 99.7065 | 99.7798 | 99.8540 | 99.7057 | 0.9963 |
| RF, MLP, SVM | 99.6800 | 99.7598 | 99.7903 | 99.7294 | 0.9963 |
| RF, kNN, SVM | 99.7171 | 99.7877 | 99.8407 | 99.7348 | 0.9966 |
| MLP, kNN, SVM | 99.6305 | 99.7226 | 99.7372 | 99.7081 | 0.9958 |

TABLE V: The results of the stacked supervised learning algorithms with ham and phishing emails. Base learners were chosen from the top performing classifiers of Table III. The meta-classifier was chosen to be Logistic Regression for all tests. All values except AUC are percentages.

| Base Learners | Accuracy | F1 | Recall | Precision | AUC |
|---|---|---|---|---|---|
| RF, MLP, kNN | 98.4496 | 98.4496 | 99.6078 | 97.3180 | 0.9846 |
| RF, MLP, SVM | 98.6434 | 98.6460 | 100.0 | 97.3282 | 0.9866 |
| RF, kNN, SVM | 98.6434 | 98.6460 | 100.0 | 97.3282 | 0.9866 |
| MLP, kNN, SVM | 98.6434 | 98.6460 | 100.0 | 97.3282 | 0.9866 |

TABLE VI: The results of spam and phishing classification using OC-SVM. The test set was balanced with half of the examples being ham and the other half being either spam or phishing.

| Metrics | Ham and Phishing | Ham and Spam |
|---|---|---|
| Accuracy | 87.2283 | 89.3339 |
| F1 | 86.3768 | 88.2430 |
| Recall | 80.9783 | 80.0555 |
| Precision | 92.5466 | 98.2960 |
| AUC | 87.2283 | 89.3339 |

TABLE VII: The best-reported results for studies using datasets A or B. Some of the cited papers here have used more than just datasets A or B, but just the reported results for those are shown. The results for this work correspond to the stacked classifiers, where the work for dataset A is the ham+spam tests, while the A+B dataset corresponds to the ham+phishing tests.

| Paper | Dataset | Accuracy | F1 | Recall | Precision | ROC AUC |
|---|---|---|---|---|---|---|
| [33] | A | - | - | - | - | 0.99 |
| [34] | A | - | - | - | - | 0.997 |
| [35] | A | 0.99 | - | - | - | - |
| [36] | A | 0.9749 | 0.9804 | - | - | 0.9951 |
| [37] | B | - | - | - | 0.98 | - |
| [38] | B | 0.9988 | 0.9946 | 0.9893 | 1.00 | |
| [40] | B | 0.9674 | 0.9671 | 0.9598 | 0.9745 | - |
| This work | A | 0.9971 | 0.9979 | 0.9984 | 0.9973 | 0.9966 |
| This work | A+B | 0.9864 | 0.9865 | 1.00 | 0.9733 | 0.9866 |

*B. Limitations and Delimitations*

One limitation of this study is that it only used spam and ham emails from a single server in a single year. During training, however, ham emails were used, so this meant that when performing testing on phishing emails, the two classes of emails (ham and phishing) were from a different server and a different year. One question that might be asked is whether the models were able to detect the phishing emails because of characteristics unique to phishing emails, or simply because they are from a different server and a different year? Additionally, many features were tailored to dataset A, and not dataset B, during the feature extraction step. One example is the timezone feature, where a value of 0 was assigned if the Date field matched the most common timezone in dataset A, otherwise, 1 was assigned. Another example is a feature that uses the Message-ID domain as a feature, where the value of 0 was assigned if the domain matched the most common domain in the Message-ID field of dataset A, otherwise, 1 was assigned. To get around these limitations, the set of utilized features was reduced when dealing with dataset B as to only include domain matching features as these are thought to generalize well since nothing specific about dataset A is used. Another limitation of this work is that many header fields were extracted, but few were fully utilized. Instead, most header fields simply corresponded to a 'missing' feature, which checked if the particular header was present or not in a given email. Ordinal and one-hot encoding techniques were used to handle issues with inconsistent email headers and missing values.

## VI. FUTURE WORK

The following are potential research opportunities identified over the course of this study, (i) creation and testing of comprehensive datasets. One of the noted limitations of this study was the use of ham emails from a single server and a single year (2007). An ideal dataset should consist of emails from multiple servers, different years, and contain ham, spam, and various types of phishing emails; (ii) developing techniques for handling email headers concept drift, which refers to how the underlying distribution of a target variable can change over time. Concept drift can occur not only over time, however, but also with location. Different servers may receive emails from some email clients more often than others, and hence the distribution of what header fields are present and which are absent can potentially vary. This was seen when looking at the distribution of email headers between datasets A and B. Not only that, but those sending spam and phishing emails may try to change their approaches over time, which is another issue; (iii) developing robust solutions against attackers who often change their strategies to get around countermeasures. A potential technique is to use models based on the Generative Adversarial Network (GAN); (iv) the use of contextual anomalies which could take into account the normal behavior of senders, such as when they typically send emails, how many, and to whom. Privacy concerns would also need to be addressed for such an approach; (v) developing a tool for the automatic processing of header fields especially those that are difficult to correctly parse; and (vi) using sentiment analysis, which is described in [49], to detect phishing emails using attributes like a sense of urgency and language style.

## VII. CONCLUSION

Previous studies on the detection and classification of anomalies in emails have, for the most part, heavily focused on the email body and subject content. In this work, the information contained in the email header is instead used in an attempt to detect anomalies in a dataset consisting of both normal and anomalous (spam and phishing) emails. Extensive feature extraction was performed and 94 features, all relating to email header content, were extracted. Feature selection was performed using permutation feature importance testing and from this, the top 30 features were extracted for use. In terms of supervised learning, the best performing classifiers were able to achieve 97% accuracy for phishing emails and 99% accuracy with spam emails, with few false positives. The most successful supervised classifiers were found to be random forest, SVM, MLP neural networks, and KNN. These results were further improved when the best performing models were combined using a stacking approach. Unsupervised learning was also tested with a one-class SVM model. This approach was able to achieve 87% accuracy with phishing emails and 89% with spam emails. The results from this work demonstrate that email headers alone, if efficiently processed, contain adequate information to make accurate determinations into whether an email is anomalous or not, especially in the supervised case.

## REFERENCES

[1] Siti Aqilah Khamis, Cik Feresa Mohd Foozy, Mohd Firdaus Ab Aziz, and Nordiana Rahim. Header based email spam detection framework using support vector machine (svm) technique. In *International conference on soft computing and data mining*, pages 57–65. Springer, 2020.

[2] Spam vs. Phishing. URL https://www.webroot.com/ca/en/resources/tips-articles/spam-vs-phishing. Last accessed on 29/07/2021.

[3] Hong Guo, Bo Jin, and Wei Qian. Analysis of email header for forensics purpose. In *2013 International Conference on Communication Systems and Network Technologies*, pages 340–344. IEEE, 2013.

[4] What is an Email Header?, . URL https://whatismyipaddress.com/email-header. Last accessed on 29/05/2021.

[5] Asaf Cidon, Lior Gavish, Itay Bleier, Nadia Korshun, Marco Schweighauser, and Alexey Tsitkin. High precision detection of business email compromise. In *28th {USENIX} Security Symposium ({USENIX} Security 19)*, pages 1291–1307, 2019.

[6] Phishing Statistics: What an Attack Costs Your Business, . URL https://www.inky.com/blog/what-a-phishing-attack-costs-your-business. Last accessed on 29/07/2021.

[7] Kathleen M Bakarich and Devon Baranek. Something phish-y is going on here: A teaching case on business email compromise. *Current Issues in Auditing*, 14(1): A1–A9, 2020.

[8] Sheng Wen, Wei Zhou, Jun Zhang, Yang Xiang, Wanlei Zhou, Weijia Jia, and Cliff C Zou. Modeling and analysis on the propagation dynamics of modern email malware. *IEEE transactions on dependable and secure computing*, 11(4):361–374, 2013.

[9] Malware Delivery via Phishing Emails is Increasing, . URL https://www.spamtitan.com/web-filtering/malware-delivery-via-phishing-emails-is-increasing/. Last accessed on 29/07/2021.

[10] Battista Biggio, Giorgio Fumera, Ignazio Pillai, and Fabio Roli. Image spam filtering by content obscuring detection. In *CEAS*, 2007.

[11] Lukas Ruff, Jacob R Kauffmann, Robert A Vandermeulen, Grégoire Montavon, Wojciech Samek, Marius Kloft, Thomas G Dietterich, and Klaus-Robert Müller. A unifying review of deep and shallow anomaly detection. *Proceedings of the IEEE*, 2021.

[12] Varun Chandola, Arindam Banerjee, and Vipin Kumar. Anomaly detection: A survey. *ACM computing surveys (CSUR)*, 41(3):1–58, 2009.

[13] Hongzhi Wang, Mohamed Jaward Bah, and Mohamed Hammad. Progress in outlier detection techniques: A survey. *IEEE Access*, 7:107964–108000, 2019.

[14] Shikha Agrawal and Jitendra Agrawal. Survey on anomaly detection using data mining techniques. *Procedia Computer Science*, 60:708–713, 2015.

[15] Asif Karim, Sami Azam, Bharanidharan Shanmugam, Krishnan Kannoorpatti, and Mamoun Alazab. A comprehensive survey for intelligent spam email detection. *IEEE Access*, 7:168261–168295, 2019.

[16] E Calò. Spf, dkim and dmarc brief explanation and best practices, 2019.

[17] NB Harikrishnan, R Vinayakumar, and KP Soman. A machine learning approach towards phishing email detection. In *Proceedings of the Anti-Phishing Pilot at ACM International Workshop on Security and Privacy Analytics (IWSPA AP)*, volume 2013, pages 455–468, 2018.

[18] Luigi Gallo, Alessio Botta, and Giorgio Ventre. Identifying threats in a large company's inbox. In *Proceedings of the 3rd ACM CoNEXT Workshop on Big DAta, Machine Learning and Artificial Intelligence for Data Communication Networks*, pages 1–7, 2019.

[19] Omar Al-Jarrah, Ismail Khater, and Basheer Al-Duwairi. Identifying potentially useful email header features for email spam filtering. In *The sixth international conference on digital society (ICDS)*, volume 30, page 140, 2012.

[20] Aziz Qaroush, Ismail M Khater, and Mahdi Washaha. Identifying spam e-mail based-on statistical header features and sender behavior. In *Proceedings of the CUBE International Information Technology Conference*, pages 771–778, 2012.

[21] Chenwei Zhang, Xiaoyan Su, Yong Hu, Zili Zhang, and Yong Deng. An evidential spam-filtering framework. *Cybernetics and Systems*, 47(6):427–444, 2016.

[22] P Kulkarni, JR Saini, and H Acharya. Effect of header-based features on accuracy of classifiers for spam email

classification. *International Journal of Advanced Computer Science and Applications*, 11(3):396–401, 2020.

[23] Chih-Chien Wang and Sheng-Yi Chen. Using header session messages to anti-spamming. *Computers & Security*, 26(5):381–390, 2007.

[24] Tim Krause, Rafael Uetz, and Tim Kretschmann. Recognizing email spam from meta data only. In *2019 IEEE Conference on Communications and Network Security (CNS)*, pages 178–186. IEEE, 2019.

[25] Wenjuan Li, Weizhi Meng, Zhiyuan Tan, and Yang Xiang. Design of multi-view based email classification for iot systems via semi-supervised learning. *Journal of Network and Computer Applications*, 128:56–63, 2019.

[26] Khoi-Nguyen Tran, Mamoun Alazab, Roderic Broadhurst, et al. Towards a feature rich model for predicting spam emails containing malicious attachments and urls. 2014.

[27] Oluyinka Aderemi Adewumi and Ayobami Andronicus Akinyelu. A hybrid firefly and support vector machine classifier for phishing email detection. *Kybernetes*, 2016.

[28] V Christina, S Karpagavalli, and G Suganya. Email spam filtering using supervised machine learning techniques. *International Journal on Computer Science and Engineering (IJCSE)*, 2(09):3126–3129, 2010.

[29] Heinz Tschabitscher. Email Headers Can Tell You About the Origin of Spam. URL https://www.lifewire.com/email-headers-spam-1166360#received-lines. Last accessed on 12/07/2021.

[30] Importance of Feature Selection in Machine Learning. URL https://www.aretove.com/importance-of-feature-selection-in-machine-learning. Last accessed on 26/07/2021.

[31] SAURAV KAUSHIK. Introduction to Feature Selection methods with an example (or how to select the right variables?). URL https://www.analyticsvidhya.com/blog/2016/12/introduction-to-feature-selection-methods-with-an-example-or-how-to-select-the-right-variables/. Last accessed on 26/07/2021.

[32] Bohdan Pavlyshenko. Using stacking approaches for machine learning models. In *2018 IEEE Second International Conference on Data Stream Mining & Processing (DSMP)*, pages 255–258. IEEE, 2018.

[33] Juan Carlos Gomez, Erik Boiy, and Marie-Francine Moens. Highly discriminative statistical features for email classification. *Knowledge and information systems*, 31(1):23–53, 2012.

[34] Dave DeBarr and Harry Wechsler. Spam detection using clustering, random forests, and active learning. In *Sixth Conference on Email and Anti-Spam. Mountain View, California*, pages 1–6. Citeseer, 2009.

[35] Neha Sattu. *A Study of Machine Learning Algorithms on Email Spam Classification*. PhD thesis, Southeast Missouri State University, 2020.

[36] Juan Carlos Gomez and Marie-Francine Moens. Pca document reconstruction for email classification. *Computational Statistics & Data Analysis*, 56(3):741–751, 2012.

[37] Tianrui Peng, Ian Harris, and Yuki Sawa. Detecting phishing attacks using natural language processing and machine learning. In *2018 ieee 12th international conference on semantic computing (icsc)*, pages 300–301. IEEE, 2018.

[38] Andre Bergholz, Jeong Ho Chang, Gerhard Paass, Frank Reichartz, and Siehyun Strobel. Improved phishing detection using model-based features. In *CEAS*, 2008.

[39] Gilchan Park and Julia M Taylor. Using syntactic features for phishing detection. *arXiv preprint arXiv:1506.00037*, 2015.

[40] Lukáš Halgaš, Ioannis Agrafiotis, and Jason RC Nurse. Catching the phish: Detecting phishing attacks using recurrent neural networks (rnns). In *International Workshop on Information Security Applications*, pages 219–233. Springer, 2019.

[41] Carlos Laorden, Xabier Ugarte-Pedrero, Igor Santos, Borja Sanz, Javier Nieves, and Pablo G Bringas. Study on the effectiveness of anomaly detection for spam filtering. *Information Sciences*, 277:421–444, 2014.

[42] M Basavaraju and Dr R Prabhakar. A novel method of spam mail detection using text based clustering approach. *International Journal of Computer Applications*, 5(4):15–25, 2010.

[43] Rafiqul Islam and Yang Xiang. Email classification using data reduction method. In *2010 5th International ICST Conference on Communications and Networking in China*, pages 1–5. IEEE, 2010.

[44] Soma Halder, Richa Tiwari, and Alan Sprague. Information extraction from spam emails using stylistic and semantic features to identify spammers. In *2011 IEEE International Conference on Information Reuse & Integration*, pages 104–107. IEEE, 2011.

[45] Anirban Chakrabarty and Sudipta Roy. An optimized k-nn classifier based on minimum spanning tree for email filtering. In *2014 2nd International Conference on Business and Information Management (Icbim)*, pages 47–52. IEEE, 2014.

[46] Ram Basnet, Srinivas Mukkamala, and Andrew H Sung. Detection of phishing attacks: A machine learning approach. In *Soft computing applications in industry*, pages 373–383. Springer, 2008.

[47] Santiago Porras, Bruno Baruque, Belén Vaquerizo, and Emilio Corchado. Clustering ensemble for spam filtering. In *International Conference on Hybrid Artificial Intelligence Systems*, pages 363–372. Springer, 2011.

[48] Ylermi Cabrera-León, Patricio García Báez, and Carmen Paz Suárez-Araujo. Self-organizing maps in the design of anti-spam filters a proposal based on thematic categories. 2016.

[49] Walaa Medhat, Ahmed Hassan, and Hoda Korashy. Sentiment analysis algorithms and applications: A survey. *Ain Shams engineering journal*, 5(4):1093–1113, 2014.